\documentclass[aps,prl,preprint,groupedaddress,showpacs]{revtex4}
\usepackage{graphicx}
\usepackage{float}
\usepackage{color}

\begin{document}

\title{Diffuse-interface model for nanopatterning induced by self-sustained ion etch masking}

\author{S. Le Roy}
\author{E. S\o nderg\aa rd}
\affiliation{Surface du Verre et Interfaces, UMR 125 CNRS/Saint-Gobain, 93303 Aubervilliers, France}
\author{I. S. Nerb\o}
\author{M. Kildemo}
\affiliation{Physics Department, Norwegian University of Science and Technology (NTNU), NO-7491 Trondheim, Norway}
\author{M. Plapp}
\email[To whom correspondence should be addressed: ]{mathis.plapp@polytechnique.fr}
\affiliation{Physique de la Mati\`{e}re Condens\'{e}e, 
\'Ecole Polytechnique, CNRS, 91128 Palaiseau, France}

\date{\today}

\begin{abstract}
We construct a simple phenomenological diffuse-interface model for
composition-induced nano\-patterning during ion sputtering of alloys.
In simulations, this model reproduces without difficulties the 
high-aspect ratio structures and tilted pillars observed in experiments. 
We investigate the time evolution of the pillar height, both by
simulations and by {\em in situ} ellipsometry. The analysis of 
the simulation results yields a good understanding of the transitions 
between different growth regimes and supports the role of segregation 
in the pattern-formation process.
\end{abstract}

\pacs{79.20.Rf, 81.16.Rf, 64.75.St}

\maketitle

The exposure of many metal and semiconductor surfaces to a homogeneous 
flux of sputtering ions leads to the spontaneous emergence of structures 
on the nanometer scale \cite{makingwave}, which can profoundly modify 
the surface properties. While this phenomenon holds great promises 
for low-cost and large-area applications in optics and electronics,
its specific use for the design of new materials and bottom-up 
manufacturing processes is at present hampered by our incomplete 
knowledge of the underlying mechanisms of self-organization. 
Especially, the dot or pillar features obtained on several 
III-V semiconductors \cite{science,Yuba2003648} or on some 
metals \cite{Rossnagel198289} are intriguing. These systems exhibit
structures of unusually high aspect ratio as well as tilted pillars
whose emergence cannot be explained by the classical 
Bradley-Harper (BH) theory \cite{BH} based on the curvature dependence
of the sputtering yield. Instead, several authors have suggested 
that these patterns arise from shadowing effects induced by composition 
variations at the surface. Two distinct mechanisms have been proposed.
In metals, the patterning was ascribed to external seeding by deposition 
of components from the surroundings of the sample \cite{metalSi}.
More recently, the phase segregation of Ga during sputtering was
pointed out as a potential source for a self-supplied etching
shield in GaSb \cite{leroyjap}, which can account for the pillar 
growth on GaSb as well as on other III-V semiconductors.

For a theoretical understanding of the composition-driven patterning
mechanisms, extensions of the BH theory have been developed that 
take into account the influence of the surface composition on the 
evolution of the morphology \cite{comporipple,Kree}. However, in 
these models the surface geometry is described in terms of a
univalued height function $h(x,y,t)$. As a consequence, tilted 
pillars (and more generally any morphology exhibiting overhangs) 
cannot be modelled. Furthermore, the development of evolution equations
for $h$ generally relies on a small-slope approximation (for a review, 
see \cite{Makeev}), which makes their application to high-aspect
ratio structures questionable. A diffuse-interface model is an 
elegant way to overcome both of these difficulties. In such models,
surfaces are represented as smooth profiles of a scalar quantity
(in our case, density) with a small but finite width $W$. This
avoids the numerical difficulties arising from the tracking of
an interface and has made such methods hugely popular in many
different areas (for reviews, see \cite{rev2,Steinbach09}).
Here, we formulate a diffuse-interface model for pillar formation
in GaSb which contains only a small number of ingredients, 
namely (i) a difference in sputtering yield between 
the two species, (ii) phase segregation of Ga, consistent 
with the equilibrium phase diagram, and (iii) diffusion of matter in
the amorphous layer created by the ion impact. Despite its
simplicity, the model is capable of producing structures that
bear a striking resemblance to those observed experimentally.
Furthermore, the time evolution of the pillar height obtained
from simulations reproduces well the experimentally observed three
different growth regimes. The simulation results allow us to
propose a refined physical interpretation of the pattern
formation process.

\begin{figure}
	\centering
	   \includegraphics{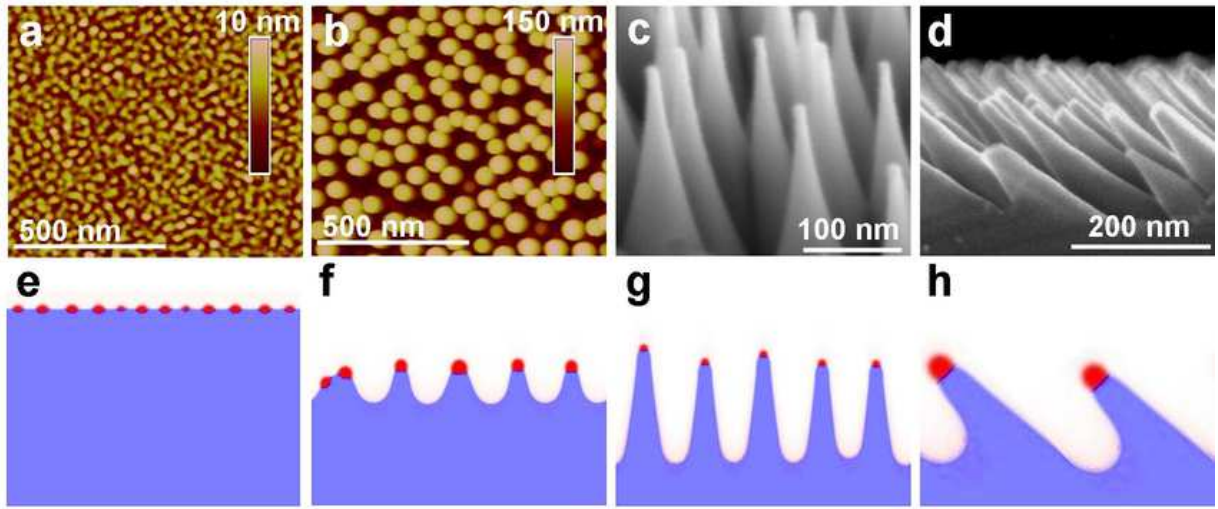}
	   \caption[figure1]{(Color online) Evolution of the surface morphology during 
sputtering from experiments and simulations: (a) and (b) AFM images of 
the surface after $30$ and $250$ s, (c) cross section SEM image after $600$ s, 
(d) cross section SEM image after $600$ s of sputtering with 45$^{\circ}$ ion 
incidence, (e), (f), (g) simulation of the surface evolution for 
increasing sputtering times, Ga is represented in red/dark gray and GaSb 
in light blue/light gray, 
(h) simulation of the surface evolution with a 45$^{\circ}$ ion incidence.}
	\label{fig:figure1}
\end{figure}

In Figure \ref{fig:figure1}a,b,c we show typical Atomic Force 
Microscopy (AFM) and Scanning Electron Microscopy (SEM) images of the
surface morphology of GaSb after exposure to a $500$ eV normal
incidence Ar$^{+}$ ion beam. In the early stage of abrasion ($30$ s), the 
surface exhibits a pattern similar to those obtained during phase
separation \cite{phaseseparation}, see Fig.~\ref{fig:figure1}a.
After $250$ s, a dense pattern of pillars has formed, Fig.~\ref{fig:figure1}b.
After $600$ s, SEM images show high-aspect-ratio features with a spherical
cap at the top. In Ref. \cite{leroyjap}, the cap was revealed to be a Ga-rich 
zone, and it was demonstrated that chemical variations precede the pattern formation.
A simultaneous segregation and shielding mechanism was proposed based
on these observations: the surface is enriched in Ga due to the difference
in sputtering yield between Ga and Sb, and above a critical concentration,
Ga segregates to form the Ga cap which acts as a sputtering shield.
Since the difference in sputtering yield as well as the tendency for
segregation are properties of the bulk material, the Ga-rich shield
is resupplied during abrasion. This mechanism can explain why
high aspect ratio structures readily form on GaSb, and also accounts
for the tilted pillars reported for oblique ion incidence (see
Fig.~\ref{fig:figure1}d and Refs.~\cite{optic,leroyjap}).

Our phenomenological model incorporates the salient 
features of the ion abrasion process, without seeking
a link to a microscopic description, which would be difficult to
achieve in view of the complexity of the ion abrasion process.
Note that our equations bear much similarity to mean-field 
kinetic equations for simple lattice-gas models that
have been used previously to describe phase separation at
surfaces \cite{spino1}.
The fundamental fields are the
dimensionless number densities (that is, the number density
times the atomic volume, assumed to be constant)
$\rho_{\rm Ga}(\vec x,t)$ and $\rho_{\rm Sb}(\vec x,t)$. The dimensionless 
free energy of the system reads
\begin{equation}
  F = \int \left[\frac 12 W^{2}\left(\nabla \rho_{\rm Ga} \right)^{2}
    + \frac 12 W^{2}\left(\nabla \rho_{\rm Sb} \right)^{2} + f_{\rm tw}\right]\;d\vec x,
\end{equation}
with
\begin{eqnarray}
 f_{\rm tw} =  &&\frac 12 \left(\rho_{\rm Ga}+\rho_{\rm Sb}\right)^{2}\left(1-\rho_{\rm Ga}-\rho_{\rm Sb}\right)^{2} + \frac 12 \left(\rho_{\rm Ga}-\rho_{\rm Sb}\right)^{2}\left(1-(\rho_{\rm Ga}-\rho_{\rm Sb})\right)^{2} \nonumber\\
  && + 8\rho_{\rm Sb}^{2}\left(0.5-\rho_{\rm Sb}\right)^{2}\;.
\end{eqnarray}
The potential $f_{\rm tw}$ has a triple-well shape, with the three
minima corresponding to the three involved phases:
$\rho_{\rm Ga}=\rho_{\rm Sb}=0.5$ (GaSb), $\rho_{\rm Ga}=\rho_{\rm Sb}=0$
(vacuum), and $\rho_{\rm Ga}=1$, $\rho_{\rm Sb}=0$ (pure Ga).
The gradient terms induce smoothing of the surfaces over a
characteristic thickness $W$. This functional generates three
two-phase equilibria: between GaSb and vacuum, between
pure Ga and vacuum, and between Ga and GaSb. Of course,
this free energy is phenomenological and differs from the
true thermodynamic free energy of the GaSb system. However,
the only feature that is fundamentally necessary for our
purpose is the room temperature coexistence of GaSb and Ga.

\begin{figure}
	\centering
	   \includegraphics{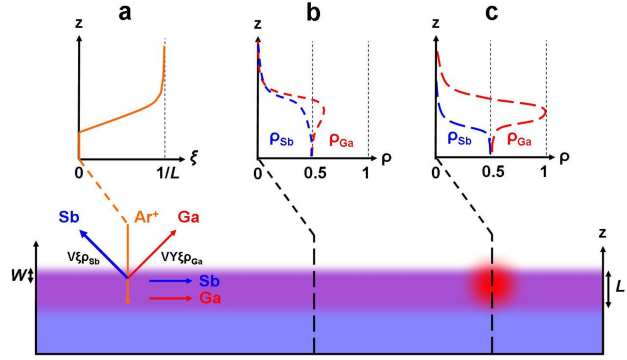}
	   \caption[figure2] {(Color online) (a) Energy distribution  $\xi$ in 
the surface layer exposed to an ion beam, leading to the ejection 
of Ga and Sb and diffusion of atomic Sb and Ga. (b),(c) density 
profiles of a surface layer before and after segregation of a gallium droplet.}
	\label{fig:shema1}
\end{figure}

Next, we introduce sputtering and diffusion. Both
processes are linked to the ion impact. Indeed, for the
temperatures prevailing in our experiments (not too far
above room temperature), bulk diffusion is virtually zero
and even surface diffusion is slow, such that substantial
mass transport occurs only due to energy transfer in the amorphous layer created
by the ion impact, see figure ~\ref{fig:shema1}a.
To account for this fact, we introduce a dimensionless quantity $\xi(\vec x,t)$
which can be seen as the fraction of the initial energy still carried by 
the impacting ion at the position $\vec x$. It satisfies the equation
\begin{equation}
\xi(\vec x,t) = {\rm max}\left(0,1 - \frac 1L \int\left(
     \rho_{\rm Ga}(\vec r (s),t)+\rho_{\rm Sb}(\vec r(s),t)\right)\;ds\right),
\end{equation}
where $L$ is the penetration depth of the ions, and $\vec r(s)$
is the ion trajectory ending at $\vec x$.
This creates an approximately linear decrease of $\xi$ in the
material, as represented in the sketch on figure \ref{fig:shema1}a.
This is a very crude approximation and differs from the usual Gaussian
shape of the deposited energy used in the theory of
Sigmund \cite{Sigmund2}. 
However, when the patterning process is composition-driven,
the precise shape of the energy distribution is not important
since the BH effect is not the principal driving phenomenon.

The evolution of the densities is then given by two equations
accounting for diffusion and abrasion,
\begin{equation}
  \frac{\partial \rho_{i}}{\partial t} =
  \vec{\nabla}\cdot\left[M_{i}(\xi,\rho_{\rm Ga},\rho_{\rm Sb}) \vec\nabla \mu_{i}\right]
       -R Y_{i} \xi\rho_{i}
  \label{eq}
\end{equation}
with $i=$ Ga or Sb,
\begin{equation}
\mu_{i} = \frac{\delta F}{\delta \rho_{i}}
\end{equation}
the local chemical potential, $R$ the sputtering rate of a flat Sb
surface (proportional to the ion flux); $Y_{\rm Sb}=1$ and $Y_{\rm Ga}=Y$
express the yield ratio between Ga and Sb. The mobilities
are given by
\begin{equation}
M_{i}(\xi,\rho_{\rm Ga},\rho_{\rm Sb}) = M_{i}^0 \xi\left(\rho_{\rm Ga}+\rho_{\rm Sb}\right)^2,
\end{equation}
where $M_{i}^0$ are constants, the factor $\xi$ accounts for the
fact that diffusion is made possible by the ion impact, and the
square of the total density suppresses spurious diffusion 
in the vacuum by evaporation and condensation,
which is negligible under the conditions of our experiments.
It is worth mentioning that, since we are using a dimensionless
free energy functional, the constants $M_{i}^0$ have the dimensions
of a diffusion coefficient.

The sputtering process is simulated by a simple finite-difference
implementation of the above equations in a box with periodic
boundary conditions on the lateral sides. The initial surface is
defined as a pure GaSb phase ($\rho_{\rm Ga}=\rho_{\rm Sb}=0.5$) in the
lower part, and vacuum in the top part ($\rho_{\rm Ga}=\rho_{\rm Sb}=0$).
A small random perturbation in both densities is added at the
GaSb-vacuum interface, which generates small fluctuations of
both surface height and surface concentration.

The parameter ranges for the quantities $W$, $M_{\rm Ga}^0$, $M_{\rm Sb}^0$, $L$, $R$, and $Y$
are selected as follows: $W$ is set to the lattice constant of
GaSb $\approx 0.6$nm, a reasonable value for the unit cell; this
parameter introduces a physical length scale in the system. The sputtering speed $RL/2$, calculated by the integration of the sputtering rate over the penetration depth, is used to introduce a time scale under the
assumption of a certain flux ($0.2\;$mA/cm$^2$) and yield (as
estimated from the simulation of the sputtering process using
the code TRIM \cite{TRIM}) which gives a sputtering speed of $0.46$ nm/s. 
$Y$ was taken between $0.4$ and $0.5$, which is small enough to obtain
phase segregation. $L$ ranges between $6$ and $12$ nm, a value close to
the estimated penetration depth \cite{TRIM}. We took 
$M_{\rm Ga}^0=M_{\rm Sb}^0=M$ for simplicity, and chose 
$50$ nm$^2$s$^{-1}<M<1000$ nm$^2$s$^{-1}$.

Figure \ref{fig:figure1}e,f,g show the time evolution 
of the surface morphology and the composition. The surface
rapidly depletes in Sb due to the difference in sputtering yield
between Ga and Sb, but remains almost flat. In this stage, the
full diffuse-interface equations can be reduced to two coupled 
equations for the surface height and the surface composition
which generalize those of Ref.~\cite{comporipple}. The details of
this procedure, as well as the linear stability analysis of
the resulting equations, will be presented elsewhere. The most
important findings are the following. For any yield ratio
(including $Y=1$) the BH instability occurs and leads to the slow 
exponential growth of structures with relatively large wavelengths.
For sufficiently small yield ratio, the surface is enriched
in Ga (Fig.~\ref{fig:shema1}b) until the interface concentration 
enters a concave region of the free energy landscape, and phase 
separation is triggered. The latter develops much faster than 
the BH instability, with a smaller characteristic wavelength, 
and leads to the formation of Ga-rich ``droplets'' 
(Fig.~\ref{fig:shema1}c) before the BH instability becomes visible.

Once Ga-rich droplets are formed, their coalescence is
prevented by the increase of the surface relief, and
well-defined structures are formed (Fig.~\ref{fig:figure1}e,f).
The mobility is a decisive parameter at this stage, since it
controls the size of and distance between the droplets. Upon
further abrasion, high aspect ratio cones are formed, with a
striking resemblance to observations by cross section SEM
(Fig.~\ref{fig:figure1}c and g). This type of time evolution 
is obtained for a wide range of parameters. The mobility can vary over
two orders of magnitude and the penetration depth from $6$ nm to $12$ nm
with no change in the pattern stability. Finally, abrasion with a
tilted ion incidence produces the same oblique structures as those
observed in experiments (Figure \ref{fig:figure1}d and h). Until now,
models have failed in predicting such structures.

\begin{figure*}
	\centering	
	\includegraphics[width=\textwidth]{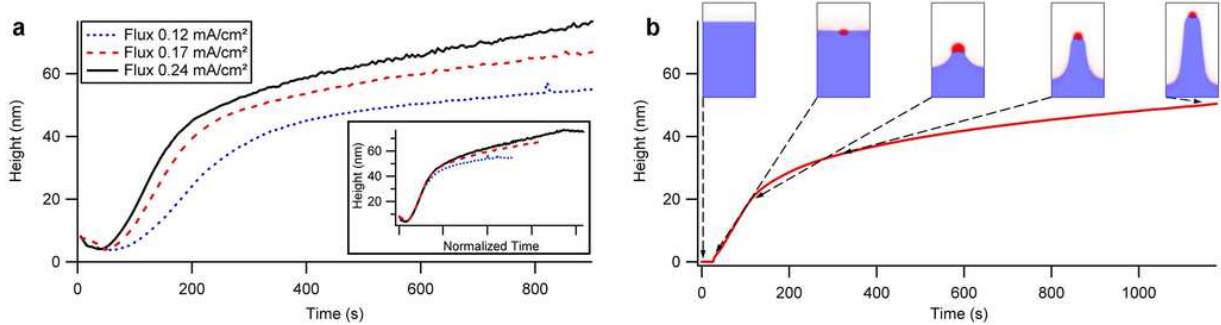}
	\caption[figure3]{(Color online) (a) Temporal evolution of the pillar height
for various flux, obtained from experiments. The inset shows curves 
where the time was normalized to the length of the initial regime, 
i.e. the time to reach a height of $7$ nm. (b) Simulated temporal 
evolution of the pillars and the corresponding morphologies, 
with $W=0.6$ nm, $L=6$ nm, $R=0.15$ s$^{-1}$, $Y=0.45$, and $M=92$ nm$^2$s$^{-1}$.}
	\label{fig:figure2}
\end{figure*}

Real-time Spectroscopic Ellipsometry (SE) measurements were performed
\textit{in situ} during the sample abrasion in
order to obtain the growth law of the features. The experiment
and data interpretation are described in Ref. \cite{optic,insitu}.
The temporal evolution of the pillar height at $300$ eV and
various fluxes is shown in figure \ref{fig:figure2}a. Three 
regimes are observed: First, a smoothing of the surface occurs.
This corresponds to the abrasion of an oxide layer initially 
present on the surface. Next, a fast growth mode follows
for around 1 minute before a slower growth mode establishes. No
saturation of the slow growth has been observed even for samples
sputtered 30 minutes \cite{insitu}.

Figure \ref{fig:figure2}b shows the simulated height evolution together
with typical surface morphologies. Since our model is phenomenological, we cannot expect
quantitative agreement with the experiments. However, we do reproduce
the three experimentally observed regimes, and the time and length
scales have the correct order of magnitude. From the simulations,
we can suggest the main mechanisms behind the three growth regimes.
The first regime corresponds to the formation of a gallium-rich and
unstable surface; the initial surface roughness is smoothed out.
In the second regime, Ga segregation has occurred and the
resulting droplets act as an etch mask. In this stage, the thickness
of the Ga droplets is approximately equal to the penetration depth.
Therefore, the growth rate is directly given by the flux multiplied
by the difference in sputtering yield between Ga and GaSb. This is 
supported by the fact that the second regime is best approximated by a
linear law rather than an exponential expression, as previously
suggested \cite{insitu}. Both in the simulation and experiment,
the slope indeed corresponds to the difference in yield between
Ga and the GaSb substrate within the uncertainty on the physical
parameters. This situation corresponds to the self-sustained etch masking
mechanism proposed in \cite{leroyjap} to explain the origin of
the nanopatterns occurring on GaSb. The time scale of the first two
regimes is directly proportional to the flux. This is evidenced from
the experimental data by the existence of a master curve when the
time is normalized by the duration of the first regime, see the
inset of Fig.~\ref{fig:figure2}a.

The transition from the second to the third regime can be understood by
analyzing the dynamic equilibrium between erosion and replenishing of
the Ga droplets. The shielding Ga droplets are maintained by a supply
of Ga from two sources. First, fresh Ga enters the droplet from
the GaSb substrate directly underneath it. Second, the GaSb
surface between the droplets continues to be enriched in Ga
by the ongoing erosion, and excess Ga diffuses along
the surface to join the droplets. However, the latter mechanism
gets weaker as the amplitude of the structures increases, because
the diffusion paths become longer. In the third regime, the diminishing Ga supply
leads to thinner droplets, which means that the Ga cap does
not completely shield against the sputtering ions. Therefore,
the contrast in the sputtering yield between shielded and unshielded
regions decreases, which leads to a slower growth of the pillars.
This partial shielding regime can last for a long time, which
explains why no saturation was found during the \textit{in situ} height
measurements. The normalized growth laws no longer fall on a master
curve (Fig.~\ref{fig:figure2}a inset). It is difficult to obtain
a general analytical expression for the growth rate in this regime,
since it is determined by an interplay between etching and
diffusion along a complex geometry, and thus depends on several
parameters in a non-trivial way.

In conclusion, the above results show that a diffuse-interface
model is a powerful method to investigate the effect of composition
variation during ion sputtering. Our simple phenomenological
model accounts for most of the specific features reported on
GaSb, such as high aspect ratio structures and tilted pillars.
The results of the simulations support the self-sustained
shielding mechanism. They also bring new insights
into the pillar formation process, and explain the overall shape
of the observed growth laws. The simulations highlight that the
limiting factor in maintaining a self-sustained etch mask is
the interplay between diffusion and a selective abrasion process.
This controls the transition from complete to partial shielding.

Clearly, the simplicity of this method makes it a promising
candidate to investigate new alleys for materials design by
spontaneous pattern formation. In conjunction with more specific
forms of the free energy as a function of composition, our approach 
can be used to investigate in a systematic way the interplay between the 
thermodynamic properties of materials and their pattern-forming ability.


\end{document}